# Evolution of high-temperature superconductivity from low-Tc phase tuned by carrier concentration in FeSe thin flakes

B. Lei[1], J. H. Cui[1], Z. J. Xiang[1], C. Shang[1], N. Z. Wang[1], G. J. Ye[1], X. G. Luo[1,4], T. Wu[1,4], Z. Sun[2,4] and X. H. Chen[1,3,4]

1. Hefei National Laboratory for Physical Sciences at Microscale and Department of Physics, University of Science and Technology of China, Hefei, Anhui 230026, China, and Key Laboratory of Strongly-coupled Quantum Matter Physics, Chinese Academy of Sciences, School of Physical Sciences, University of Science and Technology of China, Hefei, Anhui 230026, China

2. *National Synchrotron Radiation Laboratory, University of Science and Technology of China, Hefei, Anhui 230026, China*

3. High Magnetic Field Laboratory, Chinese Academy of Sciences, Hefei, Anhui 230031, China

4. Collaborative Innovation Center of Advanced Microstructures, Nanjing University, Nanjing 210093, China

**In contrast to bulk FeSe superconductor, heavily electron-doped FeSe-derived superconductors show relatively high $T_c$ without hole Fermi surfaces and nodal superconducting gap structure[1-5], which pose great challenges on pairing theories in the iron-based superconductors. In the heavily electron-doped FeSe-based superconductors, the dominant factors and the exact working mechanism that is responsible for the high $T_c$ need to be clarified [6]. In particular, a clean control of carrier concentration remains to be a challenge for revealing how superconductivity and Fermi surface topology evolves with carrier concentration in bulk FeSe. Here, we report the evolution of superconductivity in the FeSe thin flake with systematically regulated carrier concentrations by liquid-gating technique. High-temperature superconductivity at 48 K can be achieved only with electron doping tuned by gate voltage in FeSe thin flake with Tc less than 10 K. This is the first time**

**to achieve such a high temperature superconductivity in FeSe without either epitaxial interface or external pressure. It definitely proves that the simple electron-doping process is able to induce high-temperature superconductivity with Tc as high as 48 K in bulk FeSe. Intriguingly, our data also indicates that the superconductivity is suddenly changed from low-Tc phase to high-Tc phase with a Lifshitz transition at certain carrier concentration. These results help us to build a unified picture to understand the high-temperature superconductivity among all FeSe-derived superconductors and shed light on further pursuit of higher Tc in these materials.**

Heavily electron-doped FeSe-derived materials are currently the focus of researches in the field of iron-based superconductors[1-5]. These materials, including $A_xFe_{2-y}Se_2$[1-2], $Li_x(NH_2)_y(NH_3)_{1-y}Fe_2Se_2$[3], (Li,Fe)OHFeSe[4], monolayer FeSe film on $SrTiO_3$[5], share very similar electronic structures with only electron pockets near the Fermi level, and there is absent of nodal superconducting gap structure. These materials usually possess a high $T_c$ above 30K, and their unique characteristics differ from the FeAs-based superconductors and challenge current theories on a unified picture of superconductivity in iron-based superconductors[6]. In particular, the monolayer (ML) FeSe thin film on $SrTiO_3$ substrate has generated wide research interest because of its unexpected high-Tc superconductivity close to boiling temperature of liquid nitrogen (77K)[5, 7-12]. Bulk FeSe only exhibits superconductivity below 10 K [13]. Such sharp contrast on superconducting transition temperature (Tc) suggests that the artificial interface between FeSe and $SrTiO_3$ should be the key for such dramatic enhancement of Tc[14-18]. Then, a coming question is what the working mechanism of interface to enhance superconductivity is. Subsequent angle-resolved photoemission spectroscopy (ARPES) studies indicate that both electron doping and electron-phonon interaction from interface are crucial for the dramatic enhancement of Tc in monolayer FeSe thin film[7-10,16-17]. Furthermore, a common electronic structure with significant electron pocket around M point in Brillouin zone is found in FeSe-derived bulk high-Tc superconductors (such as (Li,Fe)OHFeSe, $Rb_xFe_{2-y}Se_2$)[19,20], which is similar to that of FeSe thin film. These results suggests that only sufficient electron doping is already enough to enhance Tc from 8.5 K in bulk to at least 40K. A recent ARPES result on potassium doped FeSe thin film has found that electron doping could induce superconductivity at

48 K. In order to fully understand the role of electron doping in FeSe thin film, we use liquid-gating technique to tune the electron density in FeSe thin flake without epitaxial interface. High-Tc superconductivity is successfully achieved with $T_c$ at 48K by electron doping. Our result definitely proves that simple electron-doping process is able to induce high-temperature superconductivity with Tc at least reach to 48K in FeSe without interface.

An electric-double-layer transistor (EDLT) using ionic liquids as the gate dielectrics is very efficient to tune the carrier concentration as well as the Fermi energy [21-22]. When a gate voltage ($V_g$) is applied, cations and anions in the ionic liquid move towards the sample or the gate electrode, depending on the voltage polarity, and form a Helmholtz electric double layer (EDL). This type of device works as a nanogap capacitor. The surface of the sample can be electrically modulated with the charge accumulation or depletion. To elucidate the nature of high temperature superconductivity in FeSe thin films [5,9], especially the superconductivity induced by carrier doping, we use FeSe thin flakes to fabricate an EDLT transport channel. Figure 1(a) depicts a schematic illustration of our FeSe-based EDLT. The ionic liquid DEME-TFSI was used to serve as the dielectric, which has been found to show a large capacitance at low temperature [23]. Cations and anions are DEME+ and TFSI-, respectively. Detailed device preparation procedures are described in the Methods section. Using EDLT, we systematically investigated the electrical transport properties of exfoliated single-crystalline FeSe thin flakes with typical thickness of about 10 nm. A continuously swept positive gate voltage with a scan rate of 1 mV s$^{-1}$ was applied at 220 K, exactly the same conditions where EDL has been found to form in previous works [21-22]. A typical R - $V_g$ curve is shown in Figure 1(b). The sheet resistance of the FeSe channel remains almost unchanged with Vg below 3.25 V then shows a small drop around $V_g \approx$ 3.75 V, and then increases rapidly when $V_g$ is ramped up to about 5.25 V, and eventually tends to saturate when $V_g$ approaches 6 V. About 40% enlargement in the sample resistance can be induced by $V_g$ = 6 V.

The FeSe single crystals were grown by flux method as described in Methods (for the characterization of FeSe single crystals, see Supplemental Information Section I). After introduction of carrier to the FeSe channel surface at 220 K with applying gate voltages, we measured the four-terminal sheet resistance R as a function of temperature T down to 2 K. The results are shown in Figure 2. Before the gate voltage is applied, the FeSe sample is superconducting with onset critical temperature $T_c^{onset}$ = 5.2 K. Here, $T_c^{onset}$ is defined as the intersection temperature between the linear extrapolation of the normal state and the superconducting transition, as illustrated in the inset of Figure 2. With increasing $V_g$, $T_c^{onset}$ slowly shifts to high temperature, and reaches 7.5 K (close to the $T_c$ in bulk FeSe single crystal) at $V_g$ = 4.0 V. Further increase of $V_g$ gives rise to a two-step superconducting transitions in the sample. The low superconducting transition temperature nearly does not change, while the high one increases significantly from ~ 30 K to above 40 K. Such two-step transition should originate from the inhomogeneity of the carrier distribution. With $V_g$ > 5.0 V, only the superconducting transition with Tc above 40 K can be observed and the $T_c^{onset}$ continuously increases with increasing $V_g$. Similar field-enhanced superconductivity has also been observed in the other materials [24]. The high $V_g$ also enhances the high temperature resistance and meanwhile suppress the low temperature resistance of the sample, which leads to an increase of the residual resistance ratio (RRR) value. At $V_g$ = 6 V (the maximum voltage used in our experiment), the onset critical temperature $T_c^{onset}$ increases to 45.3 K and the zero resistance is realized at 30 K. It should be addressed that the Tc at 6 V slightly depends on the sample and the highest Tc observed is 48.2 K (see Supplemental Information Fig.S3), which is higher than any other Fe-chalcogenide superconductors without an epitaxial interface. Upon applying $V_g$ > 6 V, the samples became highly unstable and can be easily damaged, possibly owing to the electrochemical reaction with sample above the threshold voltage.

To further reveal the evolution of electronic properties in the gate voltage tuning process, we measured the Hall resistance $R_{xy}$. The temperature dependence of Hall coefficient $R_H$ is shown in Figure 3(a). At $V_g$ = 0 V, $R_H$ changes sign from positive to negative at $T$ ~ 200 K, and changes sign back to positive at $T$ ~ 120 K, then increases rapidly with decreasing temperature. The evolution of $R_H$ with $T$ is consistent with the behavior of bulk FeSe except that the low-temperature non-linear behavior of $R_{xy}(B)$ in bulk crystal [25, 26] [see Supplemental Information Fig. S1(d)] is absent in our thin flake samples (Supplemental Information Figure S4(a)). The sign reversal and strong temperature dependence of $R_H$ could originate from the multi-bands electronic structure of FeSe. In the gate-voltage-induced high-$T_c$ phase ($V_g$ = 6.0 V, $T_c$ = 45.3 K), however, the temperature dependence of $R_H$ is strikingly different: $R_H$ shows negative, indicating that the dominant carriers is electron-type, and decreases gradually with decreasing temperature. We note that at $V_g$ = 6.0 V, $R_{xy}(B)$ also exhibits a linearity in the entire temperature range (Supplementary Information, Fig. S4(b)). The change of dominated charge carrier at low temperatures from hole-type in the as-exfoliated FeSe thin flakes to electron-type in the high-$T_c$ phase clearly indicates that applying positive voltage can introduce electron charge to the surface of the FeSe thin flake. Moreover, the evolution of $R_H$ - $T$ curves from as-exfoliated ($V_g$ = 0 V) to electron-doped ($V_g$ = 6.0 V) sample mimics the behavior of 5-uc FeSe films on SrTiO$_3$ substrate with annealing time prolonged from 0 h to 36 h [27]. It suggests that the annealing procedure and the gate voltage tuning process play the same role for enhancement of superconductivity in FeSe thin films. The enhancement of negative $R_H$ at low temperature at $V_g$ = 6.0 V has also been observed for $K_xFe_{2-y}Se_2$ single crystals with $T_c$ = 32 K [28] and optimally doped (Li, Fe) OHFeSe flake with $T_c$ = 43.4 K [29]. These similarities hint a possibly universal underlying physics in these heavily electron doped FeSe superconductors.

The evolution of Hall signal from hole-type to electron-type upon applying gate voltage is summarized in Fig. 3(b). Most intriguingly, an abrupt jump on the value of $R_H$ from positive to negative is observed at $V_g$ = 4.25 V, indicating a dramatic modification in the electronic structure in its vicinity induced by charge doping. Here we note that in a certain voltage range around this jump, the $R_{xy}$ ($B$) curves exhibit non-linear behavior as a characteristic of typical multi-band system (Supplementary Information, Fig. S5). For these curves, we extract $R_H$ from the slope at the high field end since it is an indication of the effective carrier density [25]. The most striking feature of the discontinuous jump on $R_H$ is in coincidence with the appearance of high-$T_c$ superconducting phase. In Figure 4, we plot the gate voltage dependence of $T_c^{onset}$ and the effective carrier density (i.e., Hall number $n_H$ = 1/$eR_H$, at T = 50 K) together. Upon applying $V_g$, $n_H$ initially decreases in accord with an electron doping process, while $T_c$ is also slowly enhanced. At $V_g$ = 4.25 V, the high-$T_c$ phase starts to show up, and $n_H$ shows a corresponding sudden sign reversal. Further ramping up of $V_g$ continues to gradually enhance $T_c$, whereas $n_H$ decreases continuously up to the maximum voltage.

The abnormal behavior observed at $V_g = 4.25$ V is highly likely to be related to an electronic transition that changes the band structure in FeSe. Numerous previous studies on FeSe confirm it to be a semimetal with both hole and electron pockets around the Fermi level, and the electronic transport properties are dominated by hole carriers [26]. With doping electrons into the energy band of FeSe in an EDLT, the effective hole population is suppressed and the dominant carriers change to electron at a critical voltage close to 4.25 V. The emergence of a high-$T_c$ superconducting transition at the same critical voltage strongly suggests a close connection between the evidently enhanced $T_c$ and the electron-dominated transport properties. We recall that in all the Fe-chalcogenide systems that host $T_c$ above 30K, including $A_xFe_{2-y}Se_2$[1-2], (Li,Fe)OHFeSe[4], and monolayer FeSe film on $SrTiO_3$[5], there are only electron pockets in their electronic structures [6-10, 16-20]. A reasonable scenario based on all the evidence is that the Fermi surface topology in the electron-doped FeSe is significantly modified at a critical band-filling, which is usually called Lifshitz transition [30]. Above the critical electron doping, hole pockets vanishe as the valance bands sink below the Fermi level, and the electronic structures are similar to those of FeSe-derived high-Tc systems mentioned above. As a result, an appreciably enhanced Tc as high as 48 K is observed. The sudden sign reversal in $R_H$ can serve as an indicative of the Lifshitz transition, yet further experimental evidence is required to unambiguously confirm the existence and details of such a transition.

The gate-voltage-induced high-$T_c$ superconductivity in FeSe achieved by surface charge accumulation in an EDLT configuration provides an exceptional example of how electrostatic doping, a "clean" doping process without any change in crystal structure, can effectively enhance the superconducting transition temperature. The optimal critical temperature we observed, Tc= 48K, is so far the highest in the parent material FeSe without any artificial interface. Electron doping thus is proved to be crucial in the optimization of Tc in Fe-chalcogenide systems, and the high Tc above 30 K in these systems is likely to be closely related to the electron pockets as suggested by our Hall data. Finally, we stress that using the EDLT method we have successfully controlled the electronic structures and superconductivity in FeSe, which is a semimetal in nature, and in the entire gating process the sample remains metallic. We believe that this technique seems to be promising in tuning a variety of materials, instead of being limited in the study of insulators and doped insulators. In addition, it is should be addressed that 48 K may not be the upper limit of

$T_c$ in this electron doping process, it is possible that $T_c$ can still reach a higher value with further doping. At current stage the further increase of doping level is prevented by the possible electrochemical reaction between sample and ionic liquid at higher gate voltage, which can damage the sample. A more effective method of introducing carrier into FeSe is required for such investigation.

## Methods

### Device fabrication

Single crystals of pristine tetragonal FeSe with typical size of $2\times2\times0.1\text{mm}^3$ were grown using a KCl–AlCl$_3$ flux method [31]. Fe and Se powders, KCl and anhydrous AlCl$_3$ were mixed in the mole ratio Fe: Se: KCl: AlCl$_3$ = 1: 0.94: 3.5: 7 and loaded into quartz tubes. The sealed quartz tubes were put into a horizontal tube furnace with one end heated to 390 ℃ and the other end at 260 ℃ for about four weeks. After dissolving the KCl/AlCl$_3$ in deionized water, shiny FeSe crystal pieces were obtained. To make our devices, we mechanically exfoliated thin flakes of FeSe from bulk single crystals using scotch tape and transferred them onto the surface of 300 nm thick SiO$_2$ insulating layer grown on a highly doped Si substrate. We selected atomically flat thin flakes with an optical microscopy and characterized the thickness by an atomic force microscopy. Thin flakes (typically 10 nm thick) were patterned into a standard Hall bar configuration and coated with Cr/Au electrodes (5/50 nm) for transport measurements using the lithography and lift-off techniques. A droplet of ionic liquid (DEME-TSFI) was applied onto the surface of the thin flakes covering both the sample and the gate electrode.

### Measurements

The measurements with varied temperature and magnetic field were performed using a Quantum Design physical property measurement system. The sheet resistance and Hall resistance were measured simultaneously using a lock-in amplifier (Stanford Research 830). A Keithley 2400 source meter supplies the gate voltage. The gate voltage was swept at 220K with a constant rate of 1 mV s$^{-1}$. Both the temperature and the gate voltage were held fixed for half an hour before cooling down to ensure a homogeneous modulation. To avoid destroying samples caused by drastic change of tension within the frozen electrolyte, we took a relatively slow cooling rate

(1K/min between 170 K and 120 K).

**Acknowledgements**

This work is supported by National Natural Science Foundation of China (NSFC), the "Strategic Priority Research Program (B)" of the Chinese Academy of Sciences, the National Basic Research Program of China (973 Program). Partial work was done in USTC center for micro- and nanoscale research and fabrication @USTC, Hefei, Anhui.

**Author contributions**

L. B. grew the single crystals and performed the resistivity, Hall coefficient, X-ray diffraction measurements with the help of J.C. Z.J.X. C.S. N.Z.W. G.Y. X.G.L., B.L. Z.J.X. T.W. Z.S. and X.H.C. analyzed the data. X.H.C. B.L. and Z.S. wrote the paper. X.H.C. conceived and coordinated the project, and is responsible for the infrastructure and project direction. All authors discussed the results and commented on the manuscript.

**Additional information**

The authors declare no competing financial interests. Correspondence and requests for materials should be addressed to chenxh@ustc.edu.cn.

# Figure captions

**Figure 1. Gate voltage modulation in the FeSe-based EDLT. (a):** A schematic illustration of the FeSe thin flake EDLT device. Ionic liquid DEME-TFSI serves as the dielectric, covering the sample and gate electrodes. (**b**): Gate voltage dependence of the resistance of a FeSe thin flake with typical thickness of about 10 nm during a continuously swept positive gate voltage at a scan rate of 1 mVs$^{-1}$. The inset shows an optical image of the EDLT device used in the present study.

**Figure 2. Temperature dependence of resistance for the FeSe thin flake at different $V_g$.** The sheet resistance was measured from 220 K to 2K with gate bias ranging from 0 to 6 V. As the gate voltage is ramped up, $T_c^{onset}$ is gradually enhanced. Eventually, the onset critical temperature $T_c^{onset}$ of the sample increases up to 45.3 K and the sample reaches to zero-resistance at 30 K with $V_g$ = 6 V. The inset is the magnified view of the region near the superconducting transition.

**Figure 3. Temperature and gate voltage dependences of Hall coefficient $R_H$ and Hall numbers $n_H$ in FeSe based DELT device**. (**a**): Temperature dependence of Hall coefficient $R$. $R_H = (R_{xy}/B)t$ was calculated by linear fit of $R_{xy}$ versus B plot from -9 T to 9 T. As $V_g$ increases to 6 V, $R_H$ changes sign from positive to negative at low temperature. (**b**): Gate voltage dependence of Hall coefficient $R_H$ and Hall number at 50 K. At $V_g$ = 4.25 V, $R_H$ shows a sudden sign reversal. The dominated charge carrier changes from hole-type to electron-type.

**Figure 4.** S**uperconducting transition temperature and $n_H$ per Fe as a function of gate voltage $V_g$.** The superconducting transition temperature shows a clear $V_g$ dependence. A transition from low-Tc phase to high-Tc phase occurs at $V_g$ = 4.25 V. At which a sudden sign reversal in $n_H$ per Fe happens, and the dominated charge carrier changes from hole-type to electron-type.

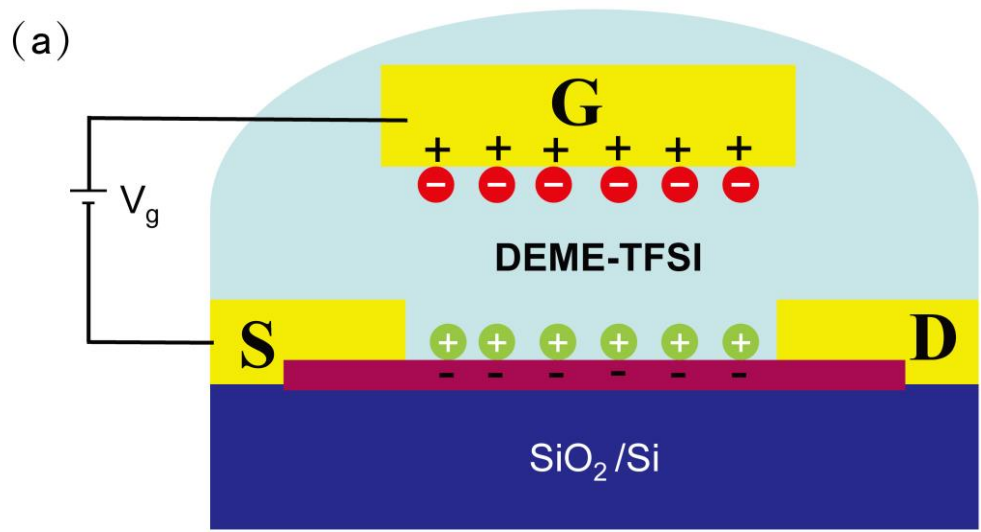

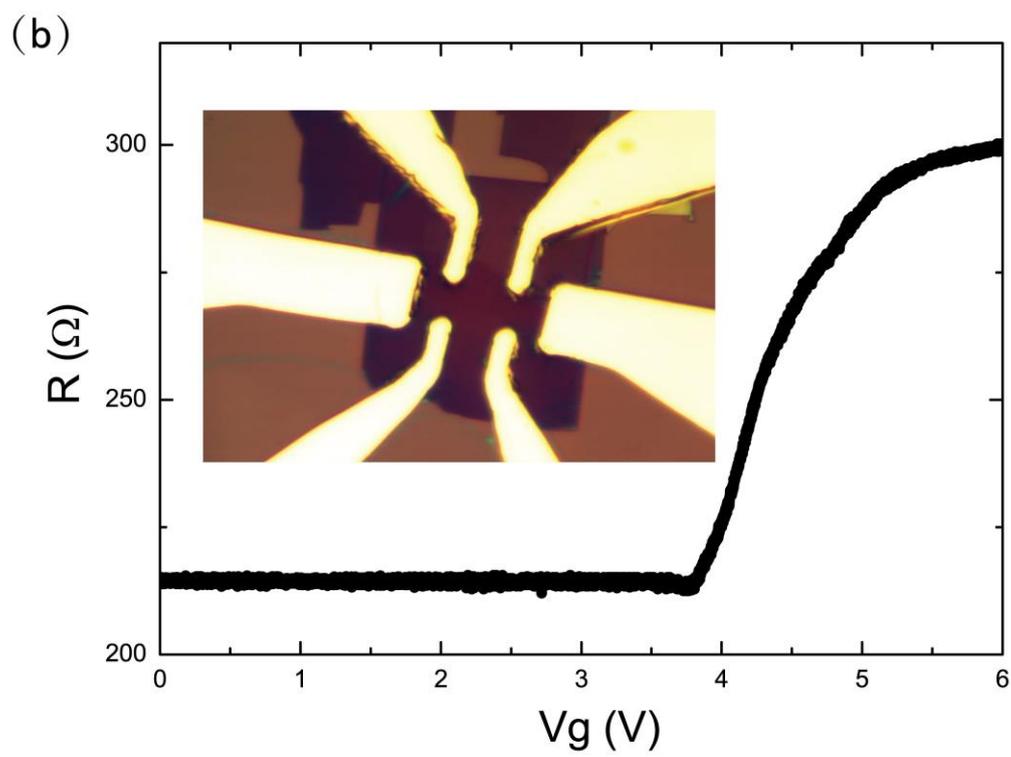

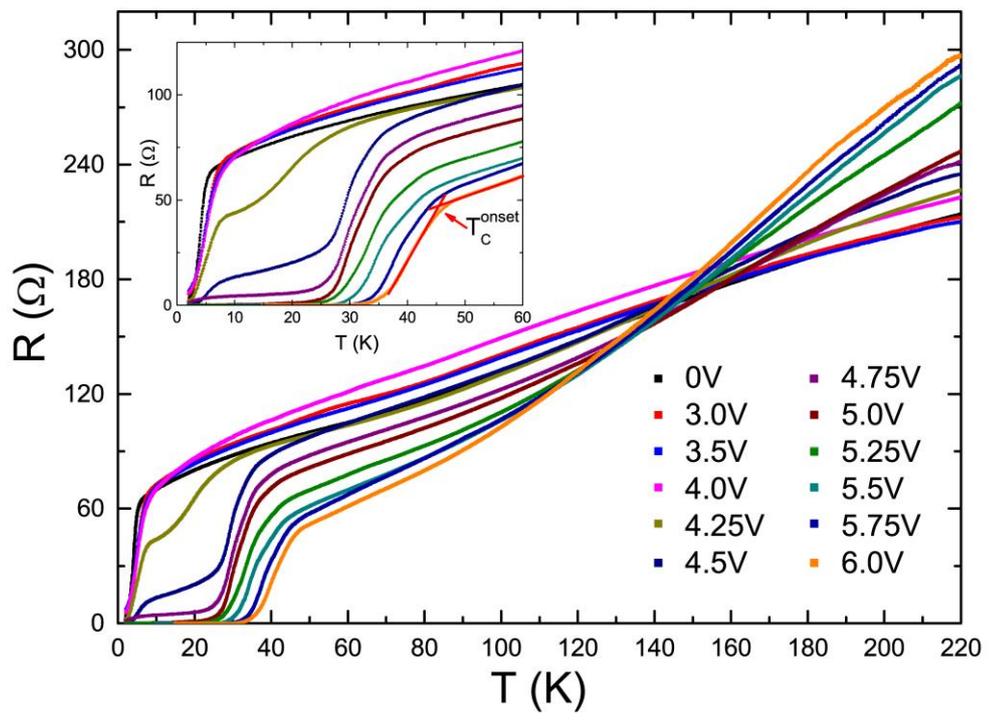

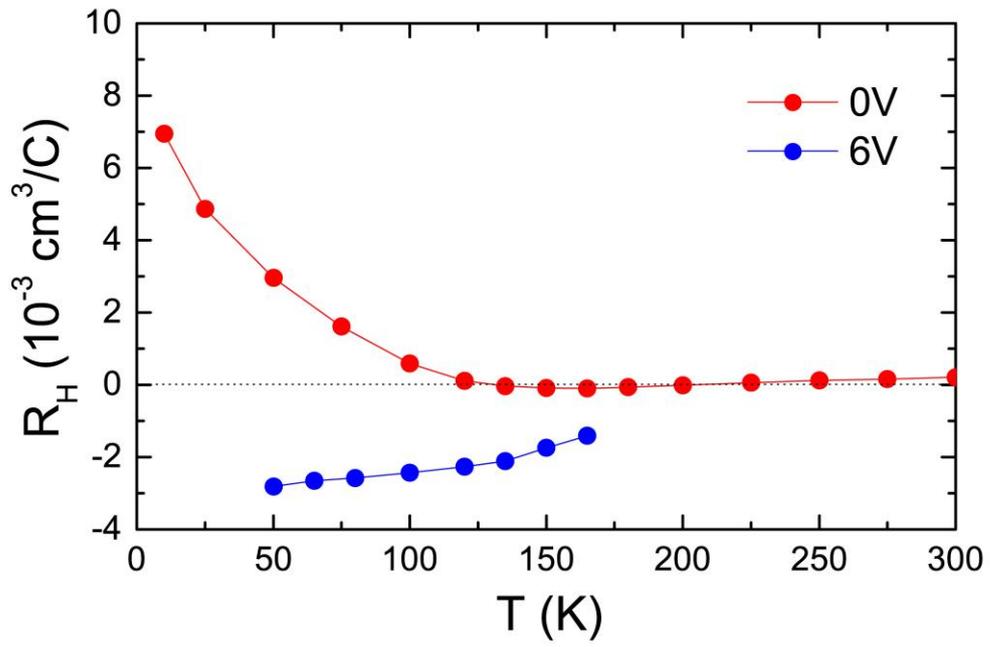
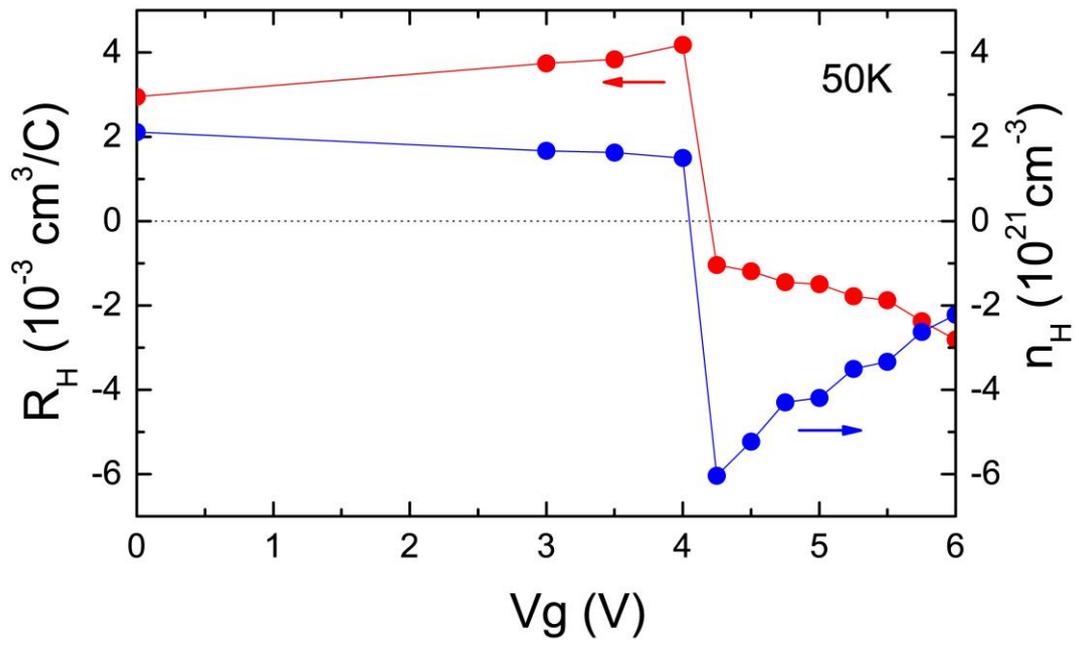

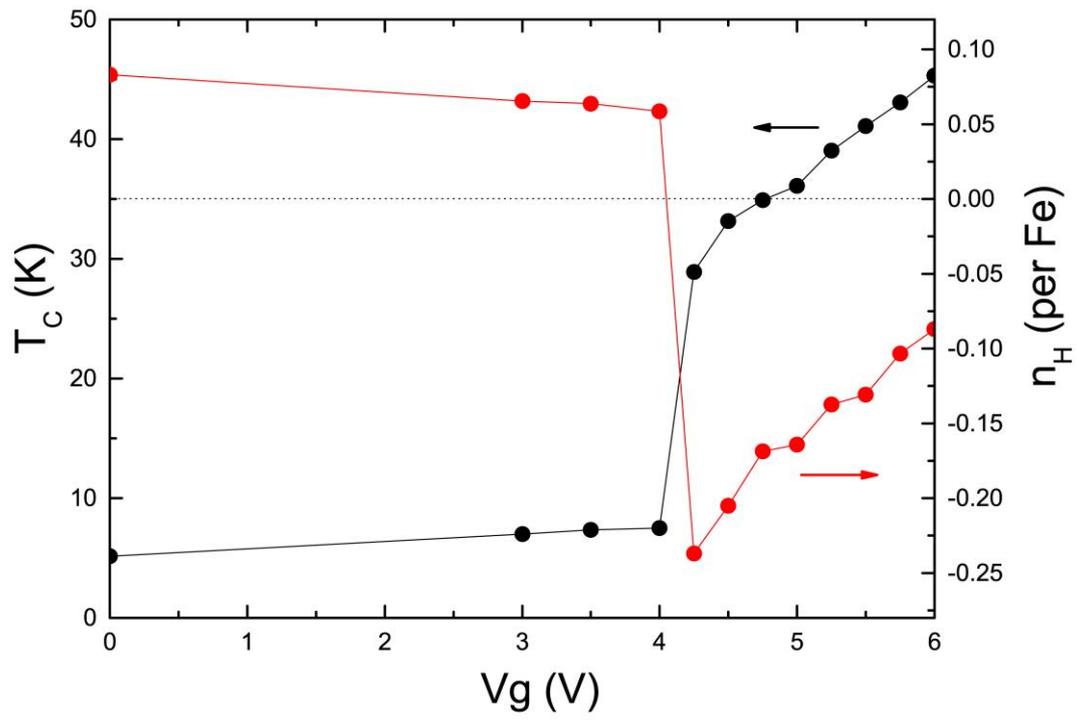